\documentclass[journal]{IEEEtai}

\usepackage[colorlinks,urlcolor=blue,linkcolor=blue,citecolor=blue]{hyperref}

\usepackage{color,array}
\usepackage{cite}
\usepackage{color,soul}
\usepackage{amsmath}
\usepackage{amsfonts}
\usepackage{multirow}
\usepackage{booktabs} 
\usepackage{graphicx}
\newcommand{\etal}{\textit{et al.}}
\begin{document}
\title{SMPL-GPTexture: Scalable, Training-Free SMPL Texture Synthesis with World Knowledge Transfer from GPT via Geometry-Aware Projection} 

\author{%
  Mingxiao Tu, Shuchang Ye, Hoijoon Jung, Jinman Kim\thanks{corresponding author} \\
  School of Computer Science\\
  University of Sydney\\
  Sydney, Australia \\
  \texttt{jinman.kim@sydney.edu.au} \\
}

\markboth{Journal of IEEE Transactions on Artificial Intelligence, Vol. 00, No. 0, Month 2025}
{\MakeLowercase{\textit{et al.}}: IEEE Journals of IEEE Transactions on Artificial Intelligence}

\maketitle

\begin{abstract}
Texturing 3D human avatars from natural-language prompts remains challenging because generated UV texture maps must be both high-quality and semantically consistent with the prompt. Prior work typically re-trains generative adversarial or diffusion models on curated datasets, which constrain generalization to clothing, stylistic, and cultural variations present in those datasets. To address these limitations, we present SMPL-GPTexture, a training-free pipeline that leverages the rich world knowledge of a pre-trained GPT model to synthesize SMPL human textures. Given a tailored natural-language prompt, GPT-4o generates geometry-consistent front/back views; we fit SMPL to each view, and invert-rasterize colours into canonical UV space with a geometry-aware view-quality weight, and close remaining gaps with diffusion-based inpainting. On a 550-prompt benchmark, SMPL-GPTexture achieves +17.6\% CLIP similarity and +22.4\% BLIP improvement over state-of-the-art text-to-texture methods while reducing visible seam artifacts.  It generalizes to diverse scenarios such as
fantasy avatars, stylized body art, and specialized uniforms without domain-specific fine-tuning. This facilitates the generation of high-quality textures that surpass current text-to-texture pipelines for fashion prototyping, AR telepresence and digital asset generation. We open-source a UV texture map dataset that covers 23 artistic styles, 49 professions, 19 ethnicities, and
67 clothing descriptors, and a set of easily adaptable prompt templates that can be repurposed to generate any desired texture, fostering research on prompt-driven avatar appearance. The code, sample dataset and prompts will be available at https://anonymous.4open.science/r/SMPL-GPTexture-2F8C.
\end{abstract}

\begin{IEEEImpStatement}
High-fidelity UV texture map generation is essential for digital fashion, augmented reality telepresence, games, and virtual production. Existing approaches require expensive multi-camera capture or costly task-specific retraining, which narrows stylistic range and limits adoption outside well-resourced studios. We present the first open-vocabulary text-to-SMPL UV texture generation framework via transferring GPT rich world knowledge, avoiding any form of task-specific data collection or training. We also release a curated set of prompt template and a benchmark of a diverse 550 UV texture maps that provide a public baseline for follow-on research or direct usage. We demonstrate that open-vocabulary UV texture generation enables the creation of diverse textures that cannot be captured in real-world settings, while showing better instruction alignment.
\end{IEEEImpStatement}


\section{Introduction}
\begin{figure}[t]
  \centering
\includegraphics[width=1\linewidth]{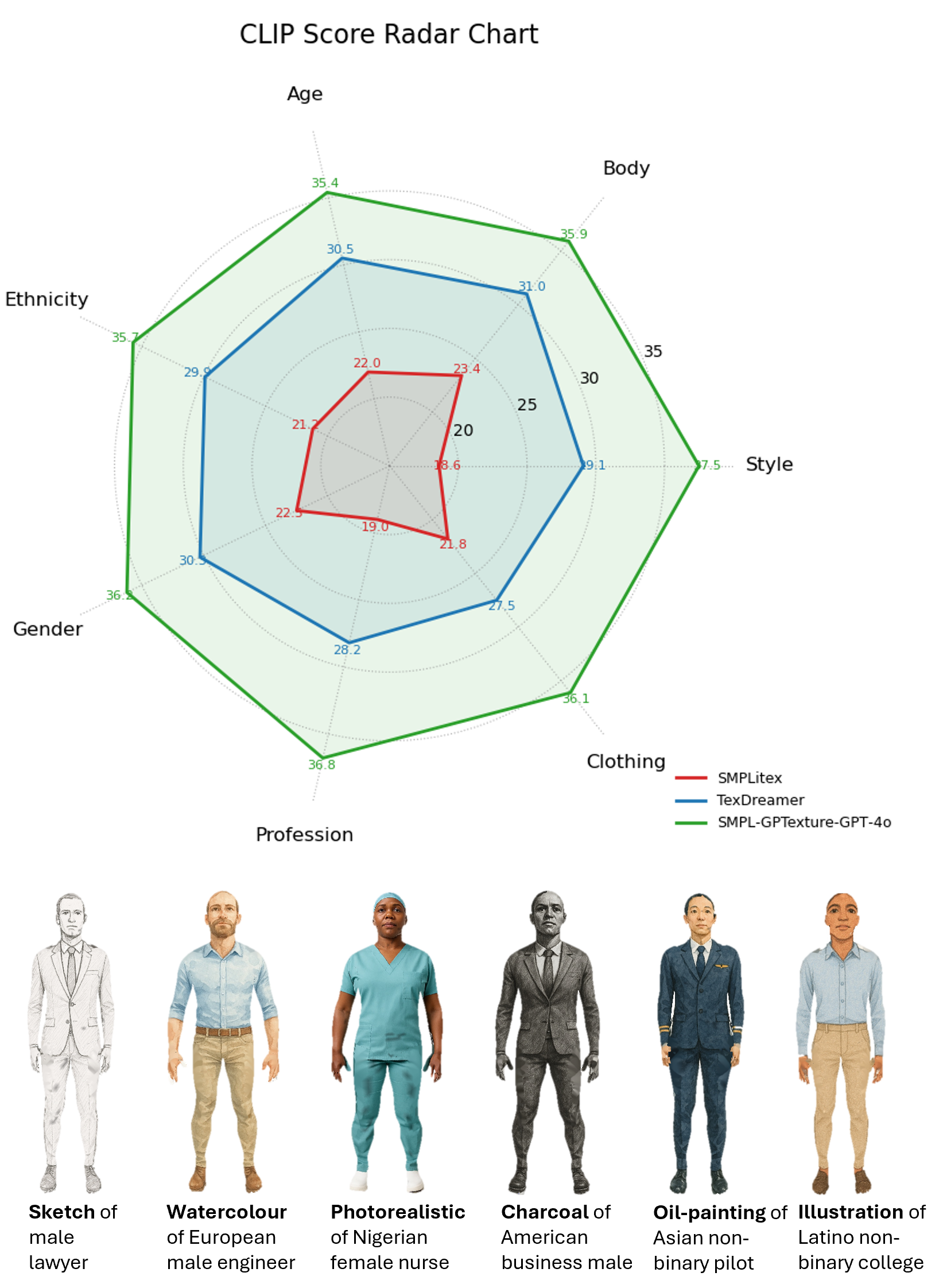}
  \caption{Performance and examples of SMPL-GPTexture. Top: CLIP score comparisons on a radar chart, evaluating SMPL-GPTexture-GPT-4o against baselines across attributes like Age, Body, Style, and Ethnicity. Bottom: A gallery of diverse, semantically coherent human textures generated from natural language prompts, demonstrating flexibility across age, gender, clothing, and artistic styles (e.g., sketch, watercolour, photorealistic).}
  \label{fig:first}
\end{figure}

\begin{figure*}[t]
  \centering
   \includegraphics[width=0.95\linewidth]{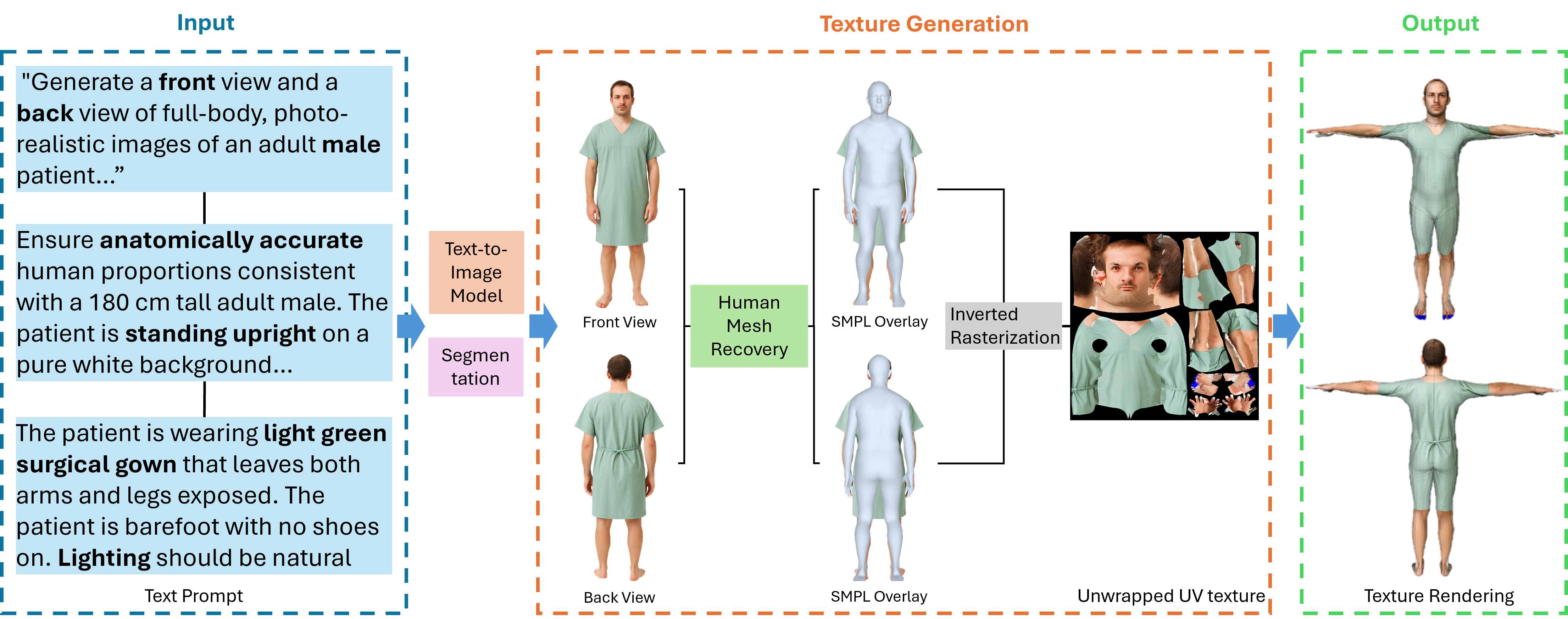}
   \caption{Overview of the proposed training-free SMPL-GPTexture pipeline. The process starts with a detailed text prompt (input), which is fed to a text-to-image model to generate consistent front and back views. A Human Mesh Recovery (HMR) model estimates the SMPL body shape from these images. Inverted rasterization is then used to project the image pixels onto an unwrapped UV texture map. This map is refined and rendered onto the 3D model to produce the final textured T-pose output (output).}
   \label{fig:model}
\end{figure*}

\IEEEPARstart{H}{igh-fidelity} texturing of 3D human avatars underpins numerous applications, including virtual try-on \cite{yang2024texture}, games \cite{zhang2023getavatar} and film production \cite{texdreamer}, e-commerce visualization \cite{He2025}, and fashion prototyping \cite{li2021toward}. In practice, components such as different skin tones, clothing, and stylistic details are encoded into a dense UV texture map, which is then mapped onto the Skinned Multi-Person Linear (SMPL) body model through UV coordinates \cite{Loper2015}. However, generating such large-scale and high-quality UV texture maps remains a challenging task. Conventional pipelines rely on 360° multi-camera rigs for acquiring human textures via calibrated RGB or photogrammetry, incurring substantial hardware cost and labour-intensive post-processing \cite{texdreamer}. As a result, existing datasets tend to be limited in size and lack stylistic variety.

To alleviate these hardware and cost constraints, image-driven methods have emerged as a scalable alternative, which synthesize the UV texture maps from single or multi-view images using deep learning models. They use convolutional inpainting and adversarial training to fill unseen regions \cite{Wang2019,Grigorev2019}, while later approaches introduced transformers to infer the occluded areas \cite{Xu2021}. Although they negate the camera-rig requirement, they still require pre-training on thousands of paired images and UV texture maps, which are scarce, and continue to struggle with artifacts in occluded regions, and are unproven for their generalizability to out-of-distribution styles and clothing \cite{texdreamer}.

To move beyond these constraints, recent work has turned to enabling UV texture generation from user text prompts by leveraging text-driven generative models. As an example, SMPLitex fine-tunes a text-to-image latent diffusion backbone on UV corpora to estimate complete 3D human textures from single images \cite{Casas2023}. Alternatively, TexDreamer employs a feature translator module to facilitate high-fidelity 3D human texture generation from text or image inputs \cite{texdreamer}. Compared with single or multi-view deep learning-based texture generation methods, these text-driven methods unlock prompt‑level control because they reuse the CLIP‑conditioned latent‑diffusion backbones and adapt them via conditional diffusion or a LoRA‑based feature‑translator to operate directly in UV space. Yet their performance remains constrained by two technical bottlenecks: (i) the available acquired training data are too small to learn a strong link between language and texture, often causing prompt–texture misalignment; and (ii) the limited visual scope of existing UV datasets hampers generalization, leading to failures on out-of-domain cases including specialized garments or highly stylized appearance such as water-colour or oil-painting effects.

In this study, we introduce SMPL-GPTexture, a training-free framework that transfers the rich world knowledge of GPT-4o \cite{openai2024gpt4o} into the texture synthesis pipeline, enabling zero-shot, on-demand generation of high-fidelity SMPL textures in arbitrary styles and outfits. The pipeline proceeds in the following three stages: first, GPT-4o produces a geometrically self-consistent pair of front- and back-view images from the tailored user prompt. Second, a lightweight human mesh recovery network is leveraged to fit an SMPL body to each portrait, establishing a precise correspondence between image pixels and mesh surface \cite{Kanazawa2018}. Third, an inverted-rasterization module projects colour from the two views onto a single UV map and a diffusion-based in-painter blends seams and fills any remaining holes, yielding a high-resolution texture ready for rendering \cite{Alldieck2018,Casas2023}. In Sec.~\ref{sec:method} we detail the pipeline components; Sec.~\ref{sec:experiment} presents the datasets, metrics and experimental protocol used for every reported evaluation; and Sec.~\ref{sec:results} provides quantitative and qualitative results, controlled ablations, and a focused discussion of limitations and failure modes (see Tables~\ref{tbl:main_results}, \ref{tbl:clip_blip_per_category}, and \ref{tab:ppshift}).

\begin{figure*}[t]
  \centering
  \includegraphics[width=0.8\linewidth]{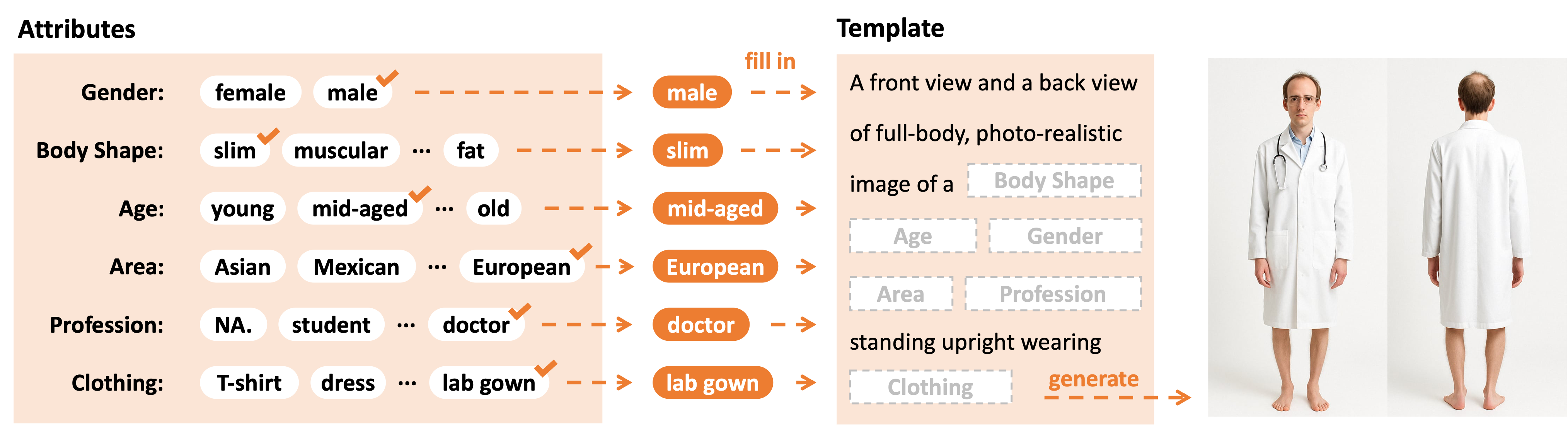}
  \caption{Structured text-to-image prompting pipeline for full-body human image generation. Specific attributes such as Gender, Body Shape, Age, Area (ethnicity), Profession, and Clothing are selected. These attributes are programmatically inserted into a detailed text template, which is then used to prompt the text-to-image model (e.g., GPT-4o) to generate consistent front and back view images.}
  \label{fig:prompt}
\end{figure*}

\begin{figure*}[t]
  \centering
  \includegraphics[width=\linewidth]{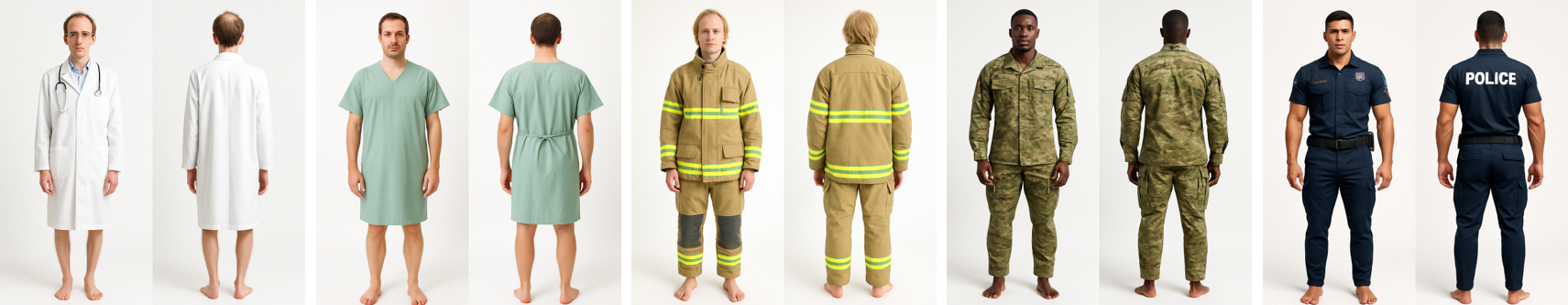}
  \caption{Raw dual-view inputs from GPT-4o used by our pipeline. This figure displays five examples of \emph{unaltered} front and back images produced by GPT-4o's structured dual-view prompting (no post-processing). These images (e.g., doctor, patient, firefighter, soldier, police officer) serve as the direct inputs for the HMR fitting and inverted rasterization stages.}
  \label{fig:gpt-generated}
\end{figure*}

Our main contributions are as follows: 

\begin{itemize}
  \item We present the first text-to-texture framework that requires no domain-specific fine-tuning yet improves generalization ability across a broad spectrum of genders, origins, ages, styles and garments, spanning photorealistic surgical scrubs, high-fashion dresses, oil-painted jerseys, and pencil-sketched uniforms, as seen in Figure~\ref{fig:first}.
  \item Using specifically crafted prompts, our synthesis step seamlessly incorporates GPT-4o image outputs and minimizes occlusion-related hallucinations when combined with inverted rasterization for UV projection and inpainting, delivering a high-quality UV texture map.
  \item We open-source a dataset of 500+ UV maps spanning 23 artistic styles, 49 professions, 19 ethnicities, and 67 clothing descriptors, fostering research on prompt-driven avatar appearance with their corresponding prompt templates, enabling reproducibility.
\end{itemize}

\noindent We evaluated our SMPL-GPTexture using in-the-wild prompts and compared it with state-of-the-art methods, demonstrating that it achieves 17.6\% higher CLIP and 22.4\% higher BLIP scores and markedly fewer seam artifacts, all without domain-specific tuning.

\section{Background}
\noindent\textbf{Conventional 3D Texture Capture.} Professional full-body photogrammetry rigs (e.g., multi-camera domes) can acquire high-fidelity 360° textures, but they are impractical for large-scale or casual use due to cost and complexity~\cite{Yu2022Humanhead,Li2022AvatarGen}. Several companies have built libraries of scanned humans with textured meshes (e.g., Renderpeople, Twindom), yet these are limited to specific captured subjects and cannot easily be adapted to new identities or imaginative designs~\cite{Li2022AvatarGen}. Moreover, the scarcity of publicly available high-quality human texture data highlights the need for automatic texture synthesis from minimal inputs~\cite{Fuentes2022Salsa}.

\noindent\textbf{Image-Driven 3D Human Texture Generation.} A number of methods estimate complete textures from one or a few input images of a person, aiming to infer 360° textures by learning priors about human appearance. Wang~\etal~\cite{Wang2019} first introduced a texture generation model supervised by person re-identification features from a single image. Texformer~\cite{Xu2021Texformer} adopted a transformer-based architecture to globally reason about visible and hidden body parts, which better estimates occluded areas than local convolutional neural networks (CNNs). Nonetheless, methods relying purely on learned single-view inference often produce blurry or incorrect textures in occluded regions~\cite{Saito2020PIFuHD}. To address this, recent approaches incorporate generative models. SiTH~\cite{Kim2024SiTH} used a diffusion model to infer a plausible occluded view conditioned on the front, then projected it onto the UV map. Similarly, ConTex-Human~\cite{gao2023contexhuman} introduced a back-view synthesis network guided by depth and a text prompt. However, aligning the generated back view with the front remains challenging. In summary, image-driven methods either require multiple views or strong priors, involve complex training pipelines, and may not generalize to wild or stylized domains.

\noindent\textbf{Text-Driven Human Texture Synthesis.} Beyond images, a recent work has explored generating textures directly from text. TexDreamer~\cite{texdreamer} fine-tuned a latent diffusion model on a curated dataset of synthetic UV-texture and prompt pairs, enabling high-resolution UV texture generation from text. However, the approach requires non-trivial retraining, and its performance is constrained by the diversity of training data. SMPLitex~\cite{Casas2023} fine-tuned a pre-trained diffusion model on a few personal textures, enabling identity-consistent outputs guided by text prompts. While effective, it requires fine-tuning for each character and does not generalize to arbitrary prompts. In general, text-driven methods either involve retraining~\cite{texdreamer,Casas2023} or per-case optimization~\cite{Kim2024SiTH,contexHuman2024}, limiting scalability and relying on a large and diversified training dataset.

\noindent\textbf{Scope of Comparison.} It is important to distinguish UV texture map generation from the broader text-to-3D human avatar generation literature. Recent methods based on 3D Gaussian Splatting (e.g., HumanGaussian~\cite{humangaussian}, GAvatar~\cite{gavatar}), deformable NeRFs (e.g., DreamHuman~\cite{dreamhuman}, DreamWaltz~\cite{dreamwaltz}), and DMTet mesh extraction (e.g., HumanNorm~\cite{humannorm}) encode appearance in per-Gaussian spherical harmonics, neural radiance fields, or arbitrary-topology meshes, respectively. These representations do not produce transferable SMPL UV texture maps and therefore address a related but distinct problem. Following the convention established by SMPLitex~\cite{Casas2023}, TexDreamer~\cite{texdreamer}, and UVMap-ID~\cite{uvmapid}, which restrict quantitative comparisons to methods that output SMPL-compatible UV maps, we adopt the same evaluation scope in this work.

Our proposed approach uses no specialized training data and needs no per-case optimization, yet it achieves high realism, prompt-faithfulness, and the ability to handle in-the-wild stylized texture with rich prior of the pre-trained world model. We now detail our approach in the next Section.

\section{Methodology}
\label{sec:method}
\subsection{System Overview}
\label{sec:overview}
SMPL-GPTexture is a training-free framework designed to synthesize high-fidelity human textures by transferring the latent world knowledge of LMMs into a geometrically constrained UV space. The pipeline follows a feed-forward progression across four modular stages: (i) \textit{Dual-View Synthesis}, which generates semantically consistent front and back portraits from structured text prompts; (ii) \textit{Structural Alignment and Validation}, where an SMPL body model is fitted to the generated images and validated for geometric consistency; (iii) \textit{Inverted Rasterization}, the core engine that lifts 2D pixel data into 3D canonical UV space; and (iv) \textit{Texture Refinement}, which employs a diffusion-based inpainter to ensure seamless UV continuity.

\subsection{Dual-View Synthesis and Geometric Validation}
\label{sec:dualview_validation}
The synthesis process begins by mapping user-defined semantic attributes, spanning gender, profession, clothing, and artistic style, into a structured prompt template. We leverage the GPT-4o to generate a high-resolution canvas containing spatially aligned front and back views $(I_{\text{front}}, I_{\text{back}})$. This approach ensures that the generated textures inherit complex visual cues (e.g., fabric textures, professional insignia) without requiring domain-specific fine-tuning.

To establish dense 3D correspondences, we fit the SMPL body model parameters $\Theta_v = \{\boldsymbol{\beta},\mathbf{p},\mathbf{t}\}$ to each view using a human mesh recovery network. To ensure structural robustness, we implement an automated rejection sampling protocol by computing the Intersection over Union (IoU) between the segmentation mask $M_v$ and the projected silhouette of the recovered mesh $\Pi_{\mathbf{K}}(\mathbf{V}_v)$ for each view $v$:
\begin{equation}
IoU = \frac{|M_v \cap \Pi_{\mathbf{K}}(\mathbf{V}_v)|}{|M_v \cup \Pi_{\mathbf{K}}(\mathbf{V}_v)|}.
\label{eq:iou}
\end{equation}
Based on empirical validation across our 550-prompt benchmark, we set a rejection threshold of $IoU < 0.75$ (flagging approximately 2.5\% of initial generations as structurally inconsistent). If a generated image fails this criterion, the system automatically triggers a re-prompting cycle. This post-generation validation effectively filters out occasional anatomical or structural inconsistencies, ensuring that only geometrically sound images proceed to the projection stage.

\begin{table}[t]
  \centering
  \caption{For each attribute, we report the number of categories (\#Cat.) present in the 550-prompt benchmark and the single most frequent example. This demonstrates the wide stylistic and demographic coverage used throughout all evaluations.}
  \label{tbl:prompt-attr}
  \begin{tabular}{lcc}
    \toprule
    \textbf{Attribute} & \textbf{\#Cat.} & \textbf{Most frequent example} \\
    \midrule
    Style                       & 23 & studio-lit \\
    Body type                   & 25 & diamond-shaped \\
    Age group                   &  6 & senior \\
    Ethnicity                   & 19 & Indigenous Australian \\
    Gender identity             &  5 & genderfluid \\
    Profession                  & 49 & student \\
    Clothing descriptor         & 67 & business casual attire \\
    \bottomrule
  \end{tabular}
\end{table}

\begin{table*}[!htbp]
  \centering
  \footnotesize
  \setlength{\tabcolsep}{3pt}
  \caption{Distribution of attributes in the 550-prompt benchmark. We list raw counts (\#) and percentages (\%) for representative groups. To demonstrate diversity, we show the top-5 categories for high-cardinality attributes; the ``others'' row aggregates the remaining long-tail distributions. Note that all columns sum to the total benchmark size ($N{=}550$).}
  \label{tbl:prompt-dist}
  \begin{tabular}{@{}lcc|lcc|lcc|lcc@{}}
    \toprule
    \multicolumn{3}{c|}{\textbf{Gender}} &
    \multicolumn{3}{c|}{\textbf{Style (top-5)}} &
    \multicolumn{3}{c|}{\textbf{Profession (top-5)}} &
    \multicolumn{3}{c}{\textbf{Clothing (top-5)}}\\
    \cmidrule{1-3}\cmidrule{4-6}\cmidrule{7-9}\cmidrule{10-12}
    Category & \# & \% &
    Category & \# & \% &
    Category & \# & \% &
    Category & \# & \% \\
    \midrule
    genderfluid         & 128 & 23.3 &
    photorealistic      &  34 &  6.2 &
    student             &  59 & 10.7 &
    business casual     &  59 & 10.7 \\
    female              & 115 & 20.9 &
    high-key            &  33 &  6.0 &
    intern              &  39 &  7.1 &
    sports jersey       &  43 &  7.8 \\
    agender             & 110 & 20.0 &
    line art            &  33 &  6.0 &
    athlete             &  37 &  6.7 &
    leotard \& tights   &  34 &  6.2 \\
    male                & 105 & 19.1 &
    gouache             &  30 &  5.5 &
    doctor              &  34 &  6.2 &
    shirt \& jeans      &  27 &  4.9 \\
    non-binary          &  92 & 16.7 &
    oil painting        &  30 &  5.5 &
    dancer              &  28 &  5.1 &
    scrubs \& badge     &  23 &  4.2 \\
    \textit{—}          & \textit{—} & \textit{—} &
    others              & 390 & 70.9 &
    others              & 353 & 64.2 &
    others              & 364 & 66.2 \\
    \bottomrule
  \end{tabular}
\end{table*}

\begin{table*}[t]
  \centering
  \caption{Mean $\pm$ standard deviation of per-prompt scores on the 550-prompt benchmark.
$^{\dag}$ denotes a statistically significant improvement over the best non-GPT-4o method
(paired two-tailed \textit{t}-test, $n{=}550$, \textit{p}$<0.01$).}
  \label{tbl:main_results}
  \setlength{\tabcolsep}{4.3pt}
  \footnotesize
  \begin{tabular}{lccc}
    \toprule
    \textbf{Method} &
    \textbf{CLIP $\uparrow$} &
    \textbf{BLIP $\uparrow$} &
    \textbf{SAS-SSIM $\uparrow$} \\
    \midrule
    SMPLitex~\cite{Casas2023} &
      $22.1 \pm 6.1$ & $14.8 \pm 4.9$ &
      $0.904 \pm 0.036$  \\[2pt]
    Paint-it~\cite{youwang2024paint} &
      $27.2 \pm 6.5$ & $20.4 \pm 5.2$ &
      $0.916 \pm 0.033$\\[2pt]
    TEXTure~\cite{richardson2023texture} &
      $28.5 \pm 6.8$ & $21.9 \pm 5.4$ &
      $0.892 \pm 0.041$ \\[2pt]
    TexDreamer~\cite{texdreamer} &
      $30.6 \pm 7.2$ & $23.7 \pm 5.8$ &
      $\mathbf{0.961 \pm 0.027} $ \\[2pt]
    SMPL-GPTexture-Infinity &
      $34.2 \pm 7.0$ & $27.5 \pm 6.0$ &
      $0.949 \pm 0.019$ \\[2pt]
    SMPL-GPTexture-SDXL & 
      $33.8 \pm 6.4$ & $27.2 \pm 5.5$ & 
      $0.951 \pm 0.018$ \\[2pt]
    SMPL-GPTexture-SD (M) &
      $35.1 \pm 5.2$ & $28.2 \pm 4.4$ &
      $0.955 \pm 0.017$ \\[2pt]
    SMPL-GPTexture-GPT-4o &
      $\mathbf{36.0 \pm 4.8^{\dag}}$ &
      $\mathbf{29.0 \pm 3.9^{\dag}}$ &
      $0.956 \pm 0.014$ \\
    \bottomrule
  \end{tabular}
  \vspace{-0.5em}
\end{table*}

\subsection{Inverted Rasterization and Texture Fusion}
\label{sec:rasterization}
Once structural alignment is validated, we populate the canonical UV space using an inverted rasterization technique. For each UV texel $(x,y)$, we solve for barycentric coordinates $(\lambda, \mu)$ to lift the texel into 3D surface coordinates $\mathbf{P}$. This point is then projected onto the input image plane using the calibrated camera parameters:
\begin{equation}
\mathbf{p}_{\text{img}} = \Pi_{\mathbf{K}}(\mathbf{R}\mathbf{P} + \mathbf{t}).
\label{eq:proj_math}
\end{equation}
To fuse the multi-view color data while minimizing artifacts at tangent surfaces, we assign a view-quality weight $w_v$ based on the dot product of the surface normal $\mathbf{n}$ and the view vector $\mathbf{v}_{\text{view}}$:
\begin{equation}
w_v(x,y) = \max(0, \mathbf{n} \cdot \mathbf{v}_{\text{view}}).
\label{eq:quality_weight}
\end{equation}
The final texture $T_{\text{final}}$ is a weighted blend of both views, preferentially sourcing pixel data from the camera with the most direct line-of-sight to the surface.

\begin{figure}[t]
\centering
\includegraphics[width=1\linewidth]{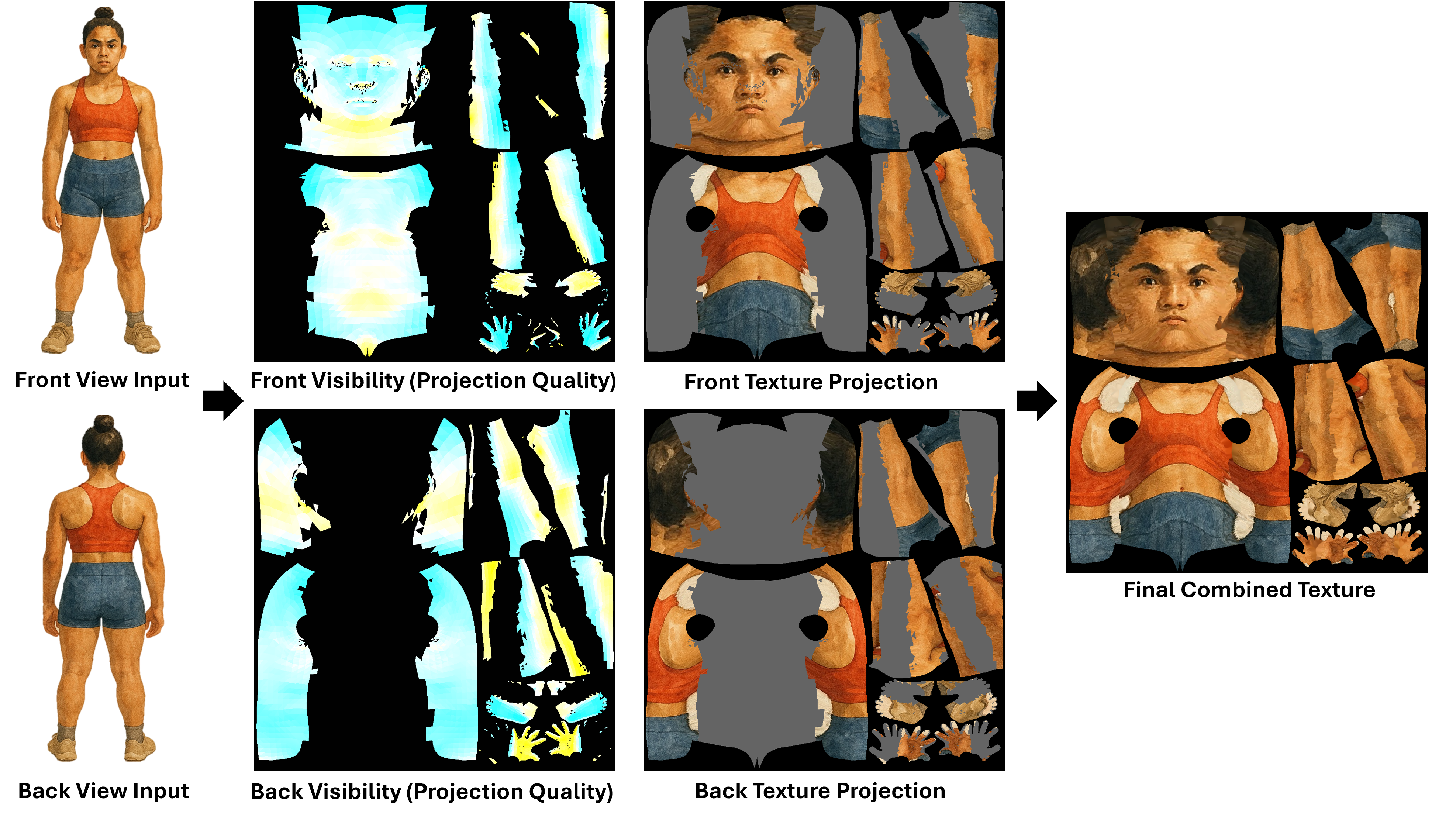}
\caption{The inverted rasterization process. Both the front and back view inputs are processed independently. For each view, a visibility mask (representing projection quality) and a partial texture projection in UV space are generated. These intermediate texture projections are then weighted by their respective visibility masks and combined to create the final, complete UV texture map, effectively blending the information from both views.}
\label{fig:inverted}
\end{figure}

\subsection{UV Completion and Refinement}
\label{sec:completion}
The raw projection may result in small uncolored regions due to self-occlusion (e.g., the lateral torso or underarms). We refine the UV map using a latent diffusion-based inpainting module. This module takes the partial UV map and the original text prompt as conditioning inputs, filling empty texels while maintaining the semantic and stylistic coherence established in the previous stages. The result is a seamless $1024 \times 1024$ UV texture map ready for standard graphics pipelines.

\section{Experimental Setup}
\label{sec:experiment}
\subsection{Implementation Details}
\label{sec:implementation}
To ensure reproducibility and isolate the effects of our proposed geometric projection, we fix all environmental and model parameters as follows.

Dual-view generation is performed via the OpenAI API using the \texttt{gpt-4o-2024-05-13} snapshot. For local diffusion-based inpainting and open-source generalizability tests, we utilize \textit{Stable Diffusion v1.5} and \textit{SDXL 1.0}. Inpainting is executed with 50 DDIM steps and a guidance scale of 7.5. Unless otherwise specified, a global random seed of 42 is used for all stochastic sampling.

The weak-perspective outputs from the HMR network are converted into a full pinhole camera model. We set the focal length $f = 1024$ px to match the generated image resolution, with the principal point at $c_x = c_y = 512$. As established in our ablation study, we introduce a fixed vertical offset $\delta_y = 8$ px to compensate for the framing bias inherent in LMM-generated portraits. All UV texture maps are synthesized at a resolution of $1024 \times 1024$ pixels.

All local computations were performed on a workstation equipped with an NVIDIA RTX-3090 GPU (24GB VRAM) and an Intel Core i9-10900K CPU. Our pipeline is implemented in PyTorch 2.1, leveraging \textit{PyTorch3D} for differentiable rendering and \textit{ScoreHMR} for body parameter regression.

\subsection{Datasets}
We evaluate on a 550-prompt benchmark constructed to stress semantic, stylistic and demographic diversity for SMPL UV generation. Prompts were created by enumerating attribute axes and instantiating a compact natural-language template that encodes: style (23 categories), body type (25), age group (6), ethnicity (19), gender identity (5), profession (49), and clothing descriptors (67). Table~\ref{tbl:prompt-attr} lists attribute counts; Table~\ref{tbl:prompt-dist} reports distributions and top categories. The prompt set intentionally mixes frequent and long-tail attribute combinations to measure robustness across realistic and stylized demands. The released artifacts include the full prompt list and the prompt template used to generate GPT-4o images for every reported experiment.

The semantic labels for this benchmark are defined procedurally; since the framework is training-free, no manual human annotation was required for dataset supervision. These procedural labels (e.g., profession and clothing type) serve as the objective ground truth for evaluating semantic alignment. To facilitate reviewer verification, we have released a subset of 50 generated examples alongside their corresponding input attributes in the supplemental repository.

\begin{figure*}
  \centering
  \includegraphics[width=0.9\linewidth]{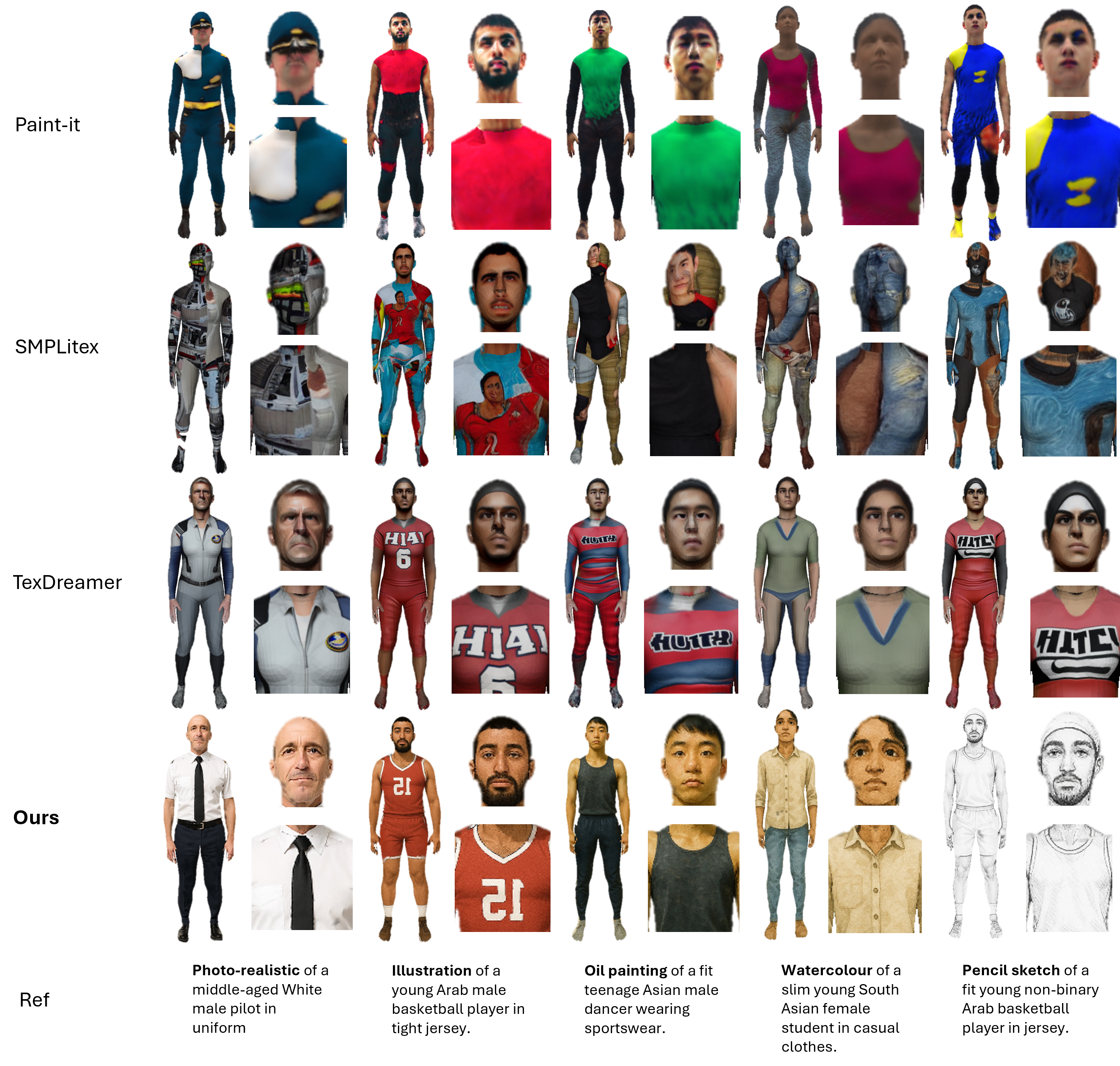}
  \caption{Qualitative comparison of texture generation in different artistic styles. Results from our method (Ours) are compared against baselines (Paint-it, SMPLitex, TexDreamer) using various text prompts as references (Ref). Our method demonstrates superior fidelity to the specified style (e.g., photorealistic, illustration, oil painting, watercolour, pencil sketch) and more accurate semantic details, such as clothing and identity cues.}
  \label{fig:style}
\end{figure*}

\begin{figure*}
  \centering
  \includegraphics[width=0.9\linewidth]{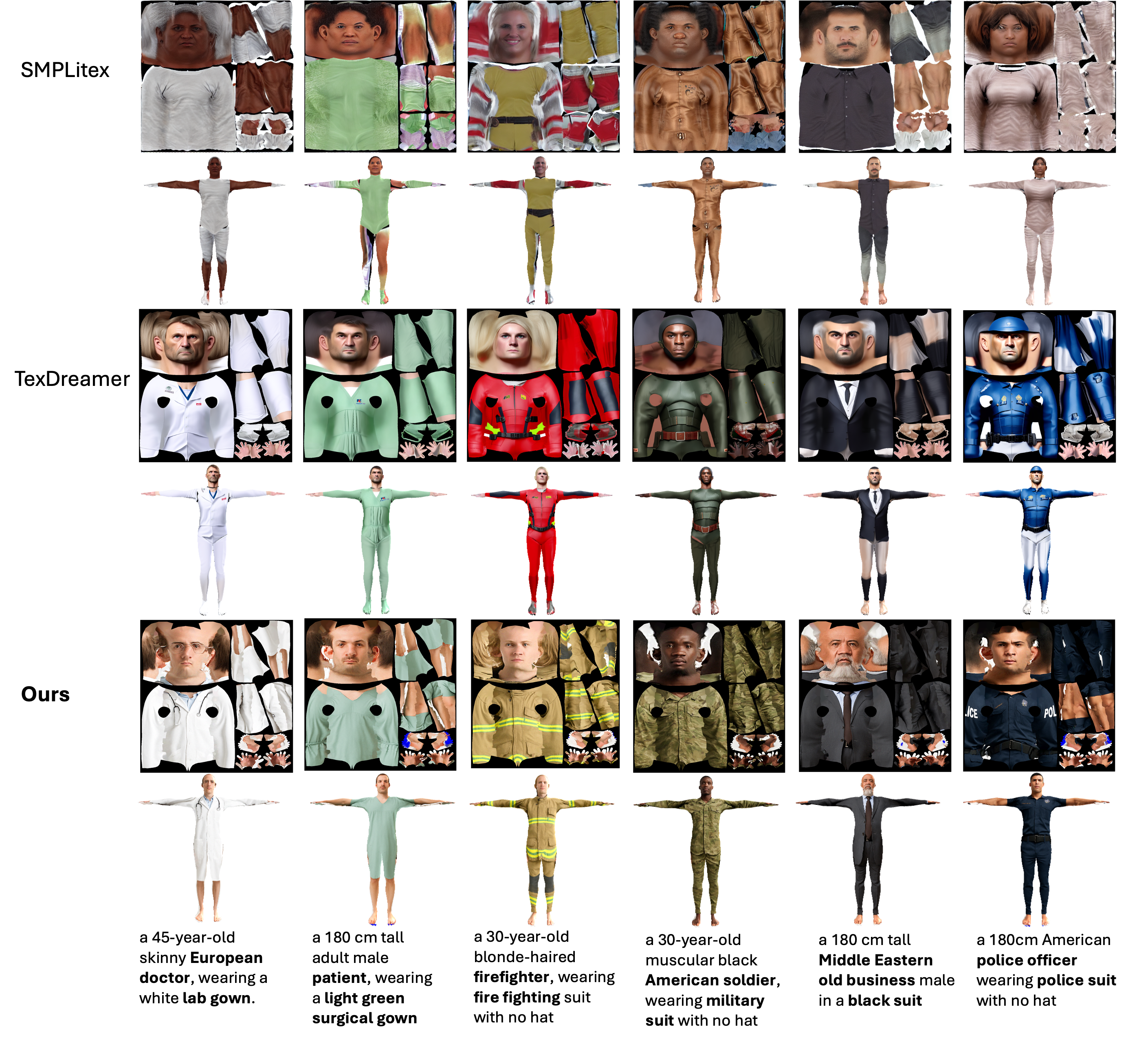}
  \caption{Qualitative results of selected texture maps generated with baselines (SMPLitex, TexDreamer) and SMPL-GPTexture (Ours) for different professions. Each method is evaluated using the same input prompt (listed at the bottom). Our method produces higher-fidelity UV maps and rendered models that more accurately reflect the detailed prompts, including specific clothing (e.g., "light green surgical gown," "black suit") and accessories.}
  \label{fig:qualitative}
\end{figure*}

\subsection{Baselines and Ablations}
We compare SMPL-GPTexture against three representative, open-source baselines that produce UV maps or textured meshes:
\begin{itemize}
  \item \textbf{SMPLitex}~\cite{Casas2023}: a diffusion-based UV estimator fine-tuned on UV corpora.
  \item \textbf{TexDreamer}~\cite{texdreamer}: a text-driven feature-translator approach for high-resolution UV maps.
  \item \textbf{Paint-It}~\cite{youwang2024paint}: a text-to-texture pipeline emphasizing stylized outputs.
  \item \textbf{TEXTure}~\cite{richardson2023texture}: an iterative optimization framework that utilizes a pre-trained depth-to-image diffusion model to progressively paint 3D surfaces from sequential viewpoints.
\end{itemize}

To isolate the contribution of the world model and geometry-aware projection we run controlled ablations while keeping downstream components fixed (same SMPL fitting, inverted rasterization, inpainting and rendering pipeline):
\begin{itemize}
  \item \textbf{SMPL-GPTexture-SD(M)} and \textbf{SMPL-GPTexture-SDXL}: replace GPT-4o dual-view generation with Stable Diffusion 3 Medium (SD3-M) and SDXL 1.0 respectively using identical prompts and sampling seeds.
  \item \textbf{SMPL-GPTexture-Infinity}: replace GPT-4o with Infinity image prior.
  \item \textbf{Vertical Alignment Ablation}: vary the vertical principal-point offset $\delta_y$ in the camera intrinsics across a range of $[-8, +8]$ pixels. This quantifies the sensitivity of the UV seam continuity to the alignment between the SMPL mesh and the GPT-generated subject.
  \item \textbf{SiTH controlled comparison}: supply SiTH~\cite{Kim2024SiTH} with the same GPT-generated front image so that differences primarily reflect algorithmic design rather than input variability.
\end{itemize}

\subsection{Evaluation metrics and protocols}
We evaluate each generated UV map using four complementary metrics and identical rendering settings for all methods:
\begin{enumerate}
  \item \textbf{CLIP similarity (prompt-to-render).} Cosine similarity between the text prompt and CLIP image embedding computed on four canonical renders (front/right/back/left; azimuths $0,90,180,270^{\circ}$) from the same SMPL pose; the per-prompt score is the mean over views and prompts (higher is better).
  \item \textbf{BLIP score (prompt–caption).} Cosine similarity between the prompt and a BLIP-2 OPT-6.7B caption embedding of the render (higher is better).
  \item \textbf{ImageReward (human-preference proxy).} A learned reward model trained on 137k human pairwise preferences~\cite{xu2023imagereward}. We evaluate front-view renders under fixed camera and material settings; a higher reward implies stronger expected human preference.
  \item \textbf{SAS-SSIM (Seam-Artifact SSIM).} A seam-focused structural similarity measure: for each UV seam, we sample a $\pm4$ px band (UV at 1024) on opposing sides, compute SSIM on $11\times11$ patches for matched texel pairs, and average across seams and prompts (higher indicates fewer seam discontinuities).
\end{enumerate}
All reported values are per-prompt averages over the 550-prompt benchmark unless otherwise specified. Per-axis averages reported in Table~\ref{tbl:clip_blip_per_category} are computed over the subset of prompts that include the corresponding attribute.
\label{sec:results}

\subsection{Reproducibility and implementation details}
We include exact settings to enable reproducibility. Unless otherwise stated, all experiments use random seed \texttt{42}. Dual-view images were generated at 1024×1024 px. GPT-4o refers to the OpenAI image-capable GPT-4o snapshot used on 2025-09-01; Infinity, SD3-M, and SDXL 1.0 indicate the public checkpoints cited in the paper. SMPL fitting uses the ScoreHMR checkpoint from~\cite{stathopoulos2024scorehmr} with default parameters; its weak-perspective outputs are converted to pinhole intrinsics as described in Section \ref{sec:implementation} and projected via Eq.~\ref{eq:proj_math}. Inpainting uses Stable Diffusion v1.5 with 50 DDIM steps and guidance scale 7.5. Geometry-aware weighting uses the surface normal dot-view term defined in Eq.~\ref{eq:quality_weight}. Rendering for metric computation uses PyTorch3D with identical camera intrinsics, lighting and material settings for all methods. Experiments were executed on nodes equipped with NVIDIA RTX-3090 GPUs. Baselines were evaluated using the same prompts, rendering and metric pipelines. The full prompt set, sample UV maps and prompts will be available at the provided repository.

\begin{table*}[t]
  \centering
  \caption{CLIP$\uparrow$ and BLIP$\uparrow$ similarity on the seven prompt axes. Each per-category cell shows \textit{CLIP (BLIP)}; best values are \textbf{bold}.}
  \label{tbl:clip_blip_per_category}
  \setlength{\tabcolsep}{4.3pt}
  \footnotesize
  \begin{tabular}{lccccccc}
    \toprule
    & \multicolumn{7}{c}{Per-category average CLIP (BLIP)} \\
    \cmidrule(lr){2-8}
    Method &
      Style & Body & Age & Ethn. & Gender & Prof. & Cloth. \\
    \midrule
    SMPLitex~\cite{Casas2023} &
      18.6\,(13.1) & 23.4\,(14.5) & 22.0\,(14.0) &
      21.2\,(13.7) & 22.5\,(14.2) & 19.0\,(13.3) & 21.8\,(14.4) \\[2pt]
    TexDreamer~\cite{texdreamer} &
      29.1\,(22.4) & 31.0\,(23.1) & 30.5\,(23.0) &
      29.9\,(22.6) & 30.3\,(22.9) & 28.2\,(21.8) & 27.5\,(21.4) \\[2pt]
    SMPL-GPTexture-SDXL & 
      34.1\,(27.4) & 33.5\,(26.8) & 33.0\,(26.3) & 
      32.6\,(26.1) & 33.7\,(27.0) & 34.0\,(27.3) & 33.8\,(27.1) \\[2pt]
    SMPL-GPTexture-GPT-4o &
      \textbf{37.5}\,(\textbf{29.7}) &
      \textbf{35.9}\,(\textbf{28.4}) &
      \textbf{35.4}\,(\textit{28.0}) &
      \textbf{35.7}\,(\textbf{28.3}) &
      \textbf{36.2}\,(\textbf{28.7}) &
      \textbf{36.8}\,(\textbf{29.2}) &
      \textbf{36.1}\,(\textbf{28.6}) \\
    \bottomrule
  \end{tabular}
\end{table*}

\begin{table}[t]
  \centering
  \footnotesize
  \setlength{\tabcolsep}{4.5pt}
  \caption{ImageReward evaluation of front-view renders across 60 random prompts. Higher reward indicates stronger human-preference alignment, while Top-1 accuracy reports the percentage of prompts where each method achieved the best score.}
  \label{tbl:imagereward}
  \begin{tabular}{lcc}
    \toprule
    \textbf{Method} & \textbf{Reward $\uparrow$} & \textbf{Top-1 Accuracy (\%) $\uparrow$} \\
    \midrule
    SMPLitex~\cite{Casas2023}    & $-0.95$ & $1.9$ \\
    Paint-It~\cite{youwang2024paint} & $-0.62$ & $3.5$ \\
    TexDreamer~\cite{texdreamer} & $-0.41$ & $5.3$ \\
    \textbf{Ours (GPT-4o)}          & $\mathbf{0.44}$ & $\mathbf{89.3}$ \\
    \bottomrule
  \end{tabular}
  \vspace{-0.5em}
\end{table}

\begin{table}[t]
\centering
\small
\setlength{\tabcolsep}{8pt}
\caption{\textbf{Ablation on vertical principal-point shift $\delta_y$ (px) (n=100).}
$\mathrm{SSIM}_{\text{seam}}$ is computed on a thin UV-boundary mask. Best in \textbf{bold}. $^{\dagger}$: significantly better than $\delta_y{=}0$ (paired two-tailed $t$-test, $p{<}.01$).}
\label{tab:ppshift}
\begin{tabular}{lc}
\toprule
$\delta_y$ (px) & $\mathrm{SSIM}_{\text{seam}}\!\uparrow$ \\
\midrule
$-8$ & $0.904 \pm 0.021$ \\
$-4$ & $0.912 \pm 0.019$ \\
$0$  & $0.915 \pm 0.018$ \\
$+4$ & $0.930 \pm 0.016$ \\
$+\mathbf{8}$ & $\mathbf{0.938 \pm 0.015}^{\dagger}$ \\
\bottomrule
\end{tabular}
\end{table}

\begin{table}[t]
  \centering
  \footnotesize
  \setlength{\tabcolsep}{6pt}
  \caption{Controlled comparison with SiTH~\cite{Kim2024SiTH} on 
  a subset of $50$ prompts. Both methods receive the same 
  GPT-generated front image; SiTH uses it directly as input, 
  while our method uses the original text prompt. 
  CLIP and BLIP are computed on four canonical rendered views.$^{*}$}
  \label{tbl:sith_quant}
  \begin{tabular}{@{}lcc@{}}
    \toprule
    \textbf{Method} & \textbf{CLIP $\uparrow$} & \textbf{BLIP $\uparrow$} \\
    \midrule
    SiTH~\cite{Kim2024SiTH} (image$\rightarrow$mesh) 
      & $31.4 \pm 6.2$ & $24.1 \pm 5.3$ \\
    \textbf{Ours} (text$\rightarrow$UV map) 
      & $\mathbf{35.8 \pm 4.9}$ & $\mathbf{28.7 \pm 4.1}$ \\
    \bottomrule
  \end{tabular}
  \vspace{0.3em}
  \par\raggedright\scriptsize{$^{*}$Note: input modalities differ 
  (image vs.\ text); SiTH produces a mesh-specific reconstruction 
  while ours produces a universal SMPL UV map. See text for details.}
  \vspace{-0.5em}
\end{table}

\begin{figure*}[t]
  \centering
  \includegraphics[width=0.8\linewidth]{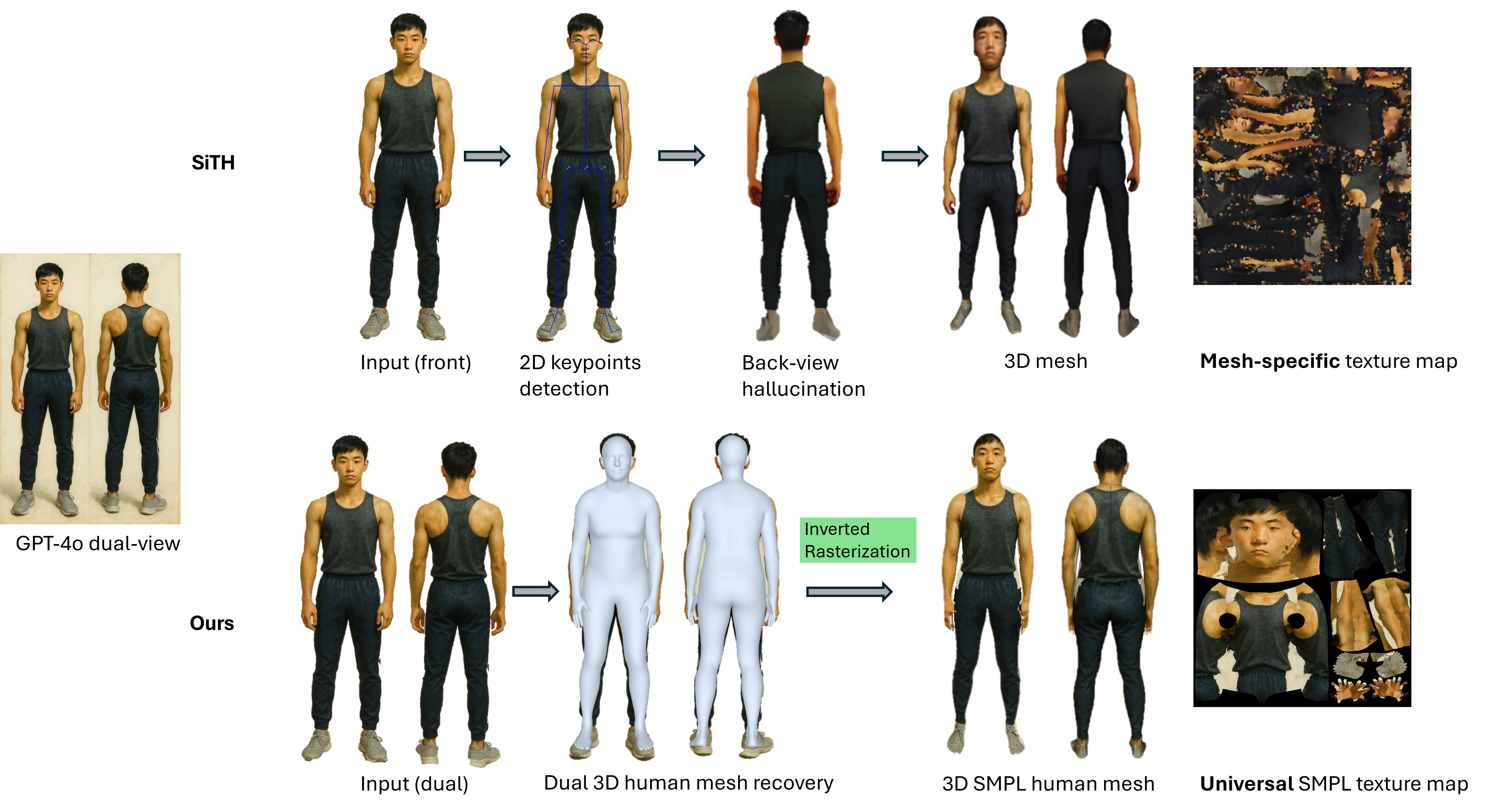}
  \caption{Ablation vs. SiTH (image$\rightarrow$texture). We feed the same GPT-generated front and back inputs to SiTH (front direct; back hallucinated per its design) and to ours. Our pipeline better preserves identity cues (face/hairline) and garment details. Importantly, our output is a canonical, \textbf{universal SMPL UV texture} that is \textbf{reusable across rigs} for downstream tasks like animation. In contrast, SiTH produces a \textbf{mesh-specific} textured mesh tied directly to its reconstruction.}
  \label{fig:ablation-sith}
\end{figure*}

\begin{figure}[t]
  \centering
  \includegraphics[width=\linewidth]{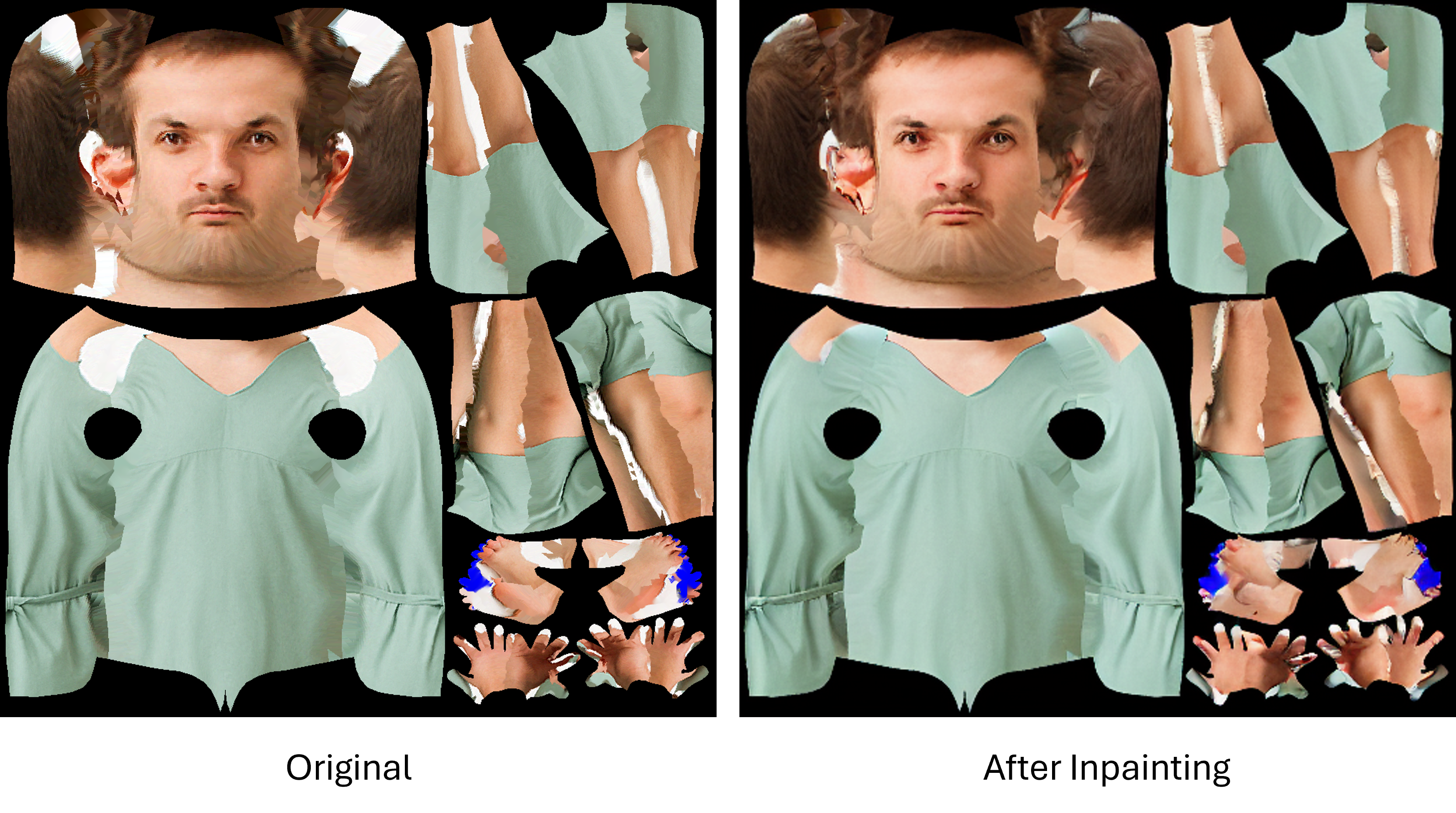}
  \caption{Effect of UV-space inpainting. We visualize the final UV texture with and without the inpainting stage. The 'Original' map (left) shows unconstrained regions (black holes) from occlusion, such as the lateral torso and underarms. 'After Inpainting' (right) shows how these regions are filled while maintaining boundary coherence, which improves SAS-SSIM by smoothing across UV cuts.}
  \label{fig:inpainting}
\end{figure}

\section{Results and Discussion}
\label{sec:results}
\subsection{Quantitative Results}
Table~\ref{tbl:main_results} reports mean ± standard deviation for CLIP, BLIP and SAS-SSIM computed across the 550-prompt benchmark; per-axis breakdowns are shown in Table~\ref{tbl:clip_blip_per_category}. Our SMPL-GPTexture-GPT-4o variant attains the highest semantic alignment with CLIP $=36.0\pm4.8$ and BLIP $=29.0\pm3.9$, improving over the strongest non-GPT variant (SMPL-GPTexture-SD(M)) by $+0.9$ CLIP and $+0.8$ BLIP (differences marked significant in Table~\ref{tbl:main_results} using paired two-tailed \textit{t}-tests at $p<0.01$). For primary comparisons (GPT-4o vs SD3-M / Infinity) typical Cohen's $d$ effect sizes for CLIP lie in the medium range ($\approx0.4$), confirming meaningful semantic gains. Furthermore, the inclusion of the open-source SDXL 1.0 backend demonstrates that our projection logic remains highly effective even when utilizing stable, non-proprietary generative models, achieving a competitive CLIP score of 33.8 while ensuring long-term reproducibility.

Per-axis analysis (Table~\ref{tbl:clip_blip_per_category}) reveals the largest GPT-specific gains on the Style and Profession axes (e.g., CLIP 37.5 vs 29.1 on Style), suggesting GPT-4o better captures stylistic and occupation-specific visual cues (logos, patterned garments, specialized textiles) that are under-represented in UV training corpora. The Infinity and SD3-M ablations reduce the gap relative to the trained baselines, indicating that improved visual fidelity alone does not fully explain GPT benefits: instead, GPT contributes richer semantic priors that the projection pipeline preserves.

Finally, we analyze the importance of geometric alignment for minimizing seam artifacts. As shown in Table~\ref{tab:ppshift}, the vertical principal-point shift $\delta_y$ impacts the SAS-SSIM score. The standard centered projection ($\delta_y=0$) yields a baseline score of $0.915 \pm 0.018$. Systematically adjusting this offset to $\delta_y=+8$ improves the seam consistency to $\mathbf{0.938 \pm 0.015}$ (a statistically significant improvement, $p < 0.01$). This indicates that GPT-generated portraits contain a slight vertical framing bias; compensating for this via $\delta_y$ is crucial for ensuring that the neck and shoulder topology aligns correctly across the UV seams.

\subsection{Qualitative Analysis}
Figures~\ref{fig:style} and \ref{fig:qualitative} show representative results across artistic styles and professions. Each panel shows (left-to-right) the prompt, GPT-4o front/back images, the fused UV map produced by inverted rasterization, and the final rendered SMPL result. Notable qualitative strengths include:
\begin{itemize}
  \item \emph{Fine-grain garment detail:} printed logos, striped patterns and reflective strips are preserved in many profession-specific examples (see firefighter, sports jersey in Fig.~\ref{fig:qualitative}).
  \item \emph{Style consistency:} painterly and line-art modalities retain characteristic strokes and tonal patterns after projection and inpainting (Fig.~\ref{fig:style}).
  \item \emph{Identity and silhouette consistency:} dual-view GPT inputs with geometry-aware projection preserve subject identity and maintain coherent silhouettes across front and back views, resulting in smoothly aligned UV maps (Fig.~\ref{fig:ablation-sith}).
\end{itemize}

To complement the qualitative comparison in  Fig.~\ref{fig:ablation-sith}, we report quantitative results in Table~\ref{tbl:sith_quant}. We evaluate on a subset of $50$ prompts using the controlled protocol described in Section~\ref{sec:experiment}: SiTH receives the GPT-generated front image directly, while our pipeline uses the original text prompt. CLIP and BLIP scores are computed on four canonical rendered views ($0^{\circ}, 90^{\circ}, 180^{\circ}, 270^{\circ}$) under identical camera and lighting settings. Three caveats apply: (i) the input modality differs, so the comparison partially reflects GPT-4o's image generation quality; (ii) SiTH outputs a mesh-specific reconstruction (marching-cubes topology) that cannot be transferred to standard SMPL rigs, unlike our universal UV map; and (iii) CLIP/BLIP measure text--render alignment, whereas SiTH is designed for image reconstruction fidelity. With these caveats, our method achieves higher semantic alignment on both metrics, consistent with the qualitative trends in Fig.~\ref{fig:ablation-sith}.

The most common failures are (i) handheld object occlusions produced by GPT-4o that remove crucial torso pixels, and (ii) extreme stylizations or poses that challenge HMR alignment near tangent regions (underarm / lateral torso). These failures are infrequent and often recoverable by re-prompting or by adding an auxiliary view.

\subsection{Efficiency Analysis}
\label{sec:efficiency}
To evaluate the practical utility of SMPL-GPTexture, we conduct a performance benchmark comparing end-to-end inference time and hardware requirements against state-of-the-art baselines. All experiments were conducted on a workstation equipped with an NVIDIA RTX-3090 GPU (24GB VRAM) and an Intel Core i9-10900K CPU.

\begin{table}[t]
  \centering
  \footnotesize
  \setlength{\tabcolsep}{6pt}
  \caption{Efficiency comparison across 100 samples. Inference time for our method includes API latency (GPT-4o) and local processing. GPU memory reflects peak usage during the inference cycle.}
  \label{tab:efficiency}
  \begin{tabular}{@{}lcc@{}}
    \toprule
    \textbf{Method} & \textbf{Time (s) $\downarrow$} & \textbf{Memory $\downarrow$} \\
    \midrule
    SMPLitex \cite{Casas2023} & $14.2 \pm 1.1$ & 6.4 GB \\
    Paint-It \cite{youwang2024paint} & $125.4 \pm 8.3$ & 12.1 GB \\
    TexDreamer \cite{texdreamer} & $18.5 \pm 2.4$ & 7.2 GB \\
    \textbf{Ours (GPT-4o)} & $\mathbf{20.8 \pm 3.5}$ & \textbf{8.5 GB} \\
    \bottomrule
  \end{tabular}
\end{table}

As shown in Table~\ref{tab:efficiency}, SMPL-GPTexture generates a high-fidelity UV map in approximately 21 seconds. While the GPT-4o API introduces a small overhead compared to single-pass models like SMPLitex, our pipeline remains significantly more efficient than optimization-based methods like Paint-it. Notably, our GPU memory footprint (8.5 GB) is well within the capacity of consumer-grade hardware, making the framework accessible for individual creators.

The reported inference time in this Table also accounts for the automated rejection loop. Since only 2.5\% of generations trigger re-prompting, the amortized overhead across the benchmark is less than 0.5 seconds per prompt on average, and the maximum observed delay for a single re-prompting cycle is bounded by one additional API call (~10–15s).

\subsection{Human-Preference Proxy Score}
We assess perceptual preference using ImageReward on front-view renders for a random subset of 60 prompts (protocol: each method renders front-view images under identical camera and material settings; ImageReward scores are averaged per prompt). Results in Table~\ref{tbl:imagereward} show SMPL-GPTexture-GPT-4o attains a mean reward $0.44$ and Top-1 accuracy $89.3\%$, substantially outperforming baselines. The large margin indicates that GPT-driven dual views substantially improve human-perceived realism, capturing garment fidelity, texture detail, and identity cues that embedding-based metrics only partially reflect.

\subsection{Prompt Robustness}
\label{sec:robustness}
We measure robustness to prompt phrasing and sampling stochasticity using the structured chat template described in Section \ref{sec:dualview_validation}. regarding sampling stochasticity, we allow up to three automated re-generation cycles based on our $IoU$ verification metric. We report \textsc{success@k} (criteria: automated pass of $IoU \geq 0.75$; instruction following; consistent dual-view alignment) on the 550-prompt set: \textsc{success@1/2/3} = 97.5\% / 98.7\% / 99.3\%. The automated rejection loop significantly reduces failures compared to a single-pass approach.

To further quantify robustness against linguistic variations, we conducted a sensitivity analysis following the text-to-image protocol of Lee et al.~\cite{lee2023holistic}. We generated semantically equivalent rephrasings for 10 base attribute sets (e.g., swapping synonyms) and computed the CLIP-space cosine similarity between textures generated from the original and rephrased prompts. Our method achieves an average similarity of $0.91 \pm 0.04$, confirming that the pipeline effectively internalizes underlying semantic concepts rather than over-fitting to specific lexical tokens.

\section{Downstream Applications}
\label{sec:downstream}
\begin{figure}[h!]
  \centering
  \includegraphics[width=0.8\linewidth]{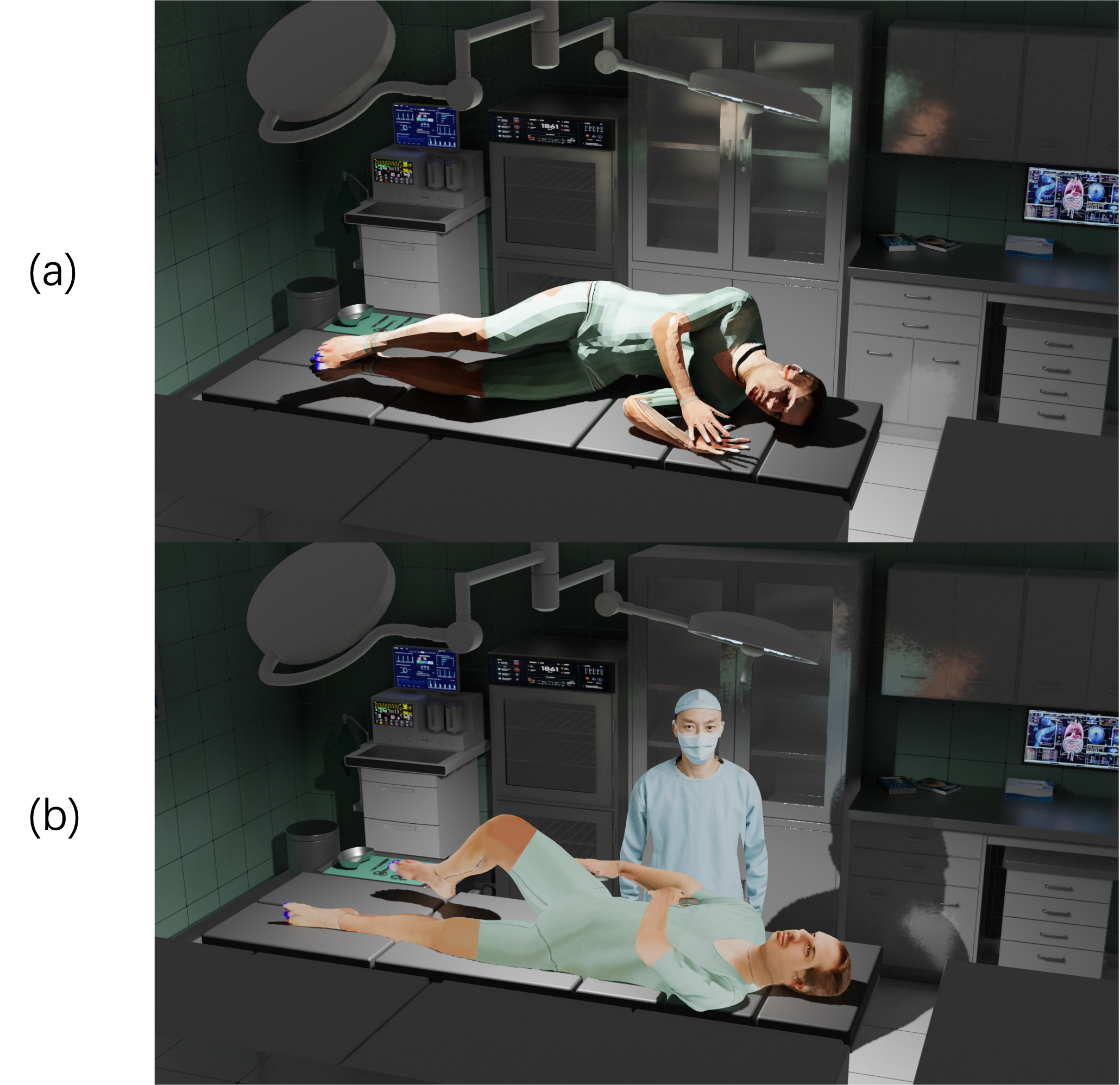}
  \caption{Digital twin simulation of a patient (textured using SMPL-GPTexture) lying on a surgical table within a 3D medical environment. (a) Simulated patient in the lateral position (b) Simulated patient for knee surgery with surgeon besides. These showcase the utility of our generated UV maps for high-fidelity scene reconstruction, enabling applications like synthetic data generation for 3D human pose estimation in clinical settings without privacy concerns.}
  \label{fig:digital_twin}
\end{figure}

We also demonstrate the potential of SMPL-GPTexture for critical downstream applications, such as generating high-fidelity digital twins for complex, privacy-sensitive medical environments.                                                        
Traditionally, acquiring high-fidelity textures of actual patients for such simulations presents significant challenges due to privacy concerns and the practical difficulties of 3D scanning in medical environments. Our SMPL-GPTexture pipeline bypasses these limitations by generating realistic patient textures solely from textual prompts. For instance, the patient texture applied in this simulation (Figure~\ref{fig:digital_twin}) is the same 'light green surgical gown' texture generated by our pipeline and depicted in Figure~\ref{fig:gpt-generated}, second person from the left.

As illustrated in Figure~\ref{fig:digital_twin}, we applied the generated patient texture to an SMPL body model and integrated it into a detailed 3D mesh of a surgical room environment. This digital twin setup allows for the generation of scalable training datasets with perfectly aligned ground truth information for a variety of AI-driven medical applications, such as 3D human pose estimation for surgical workflow analysis, patient monitoring, and ergonomics studies. The ability to create such realistic yet synthetic data addresses critical bottlenecks in medical AI research, where real-world data collection is often constrained by ethical, logistical, and privacy considerations. 

\section{Limitations and Future Work}
\label{sec:limitations}
Three systematic limitations remain. First, GPT-4o occasionally introduces hand-held objects or overlapping views, which occlude torso texels. These visual details are often "hallucinations", plausible but geometrically inconsistent elements common in LMMs, as noted in recent surveys \cite{yin2024survey}. To ensure robustness, we implement a self-correction mechanism; our $IoU \geq 0.75$ threshold (see Eq.~\ref{eq:iou}) effectively discards these structurally compromised samples before projection. 

Second, dual-view inputs under-constrain tangent surfaces, leading to occasional mesh discontinuities at side regions. While diffusion-based inpainting currently mitigates these gaps (see Fig.~\ref{fig:inpainting}), a concrete path for future improvement lies in extending the framework to a four-view synthesis strategy ($0^{\circ}, 90^{\circ}, 180^{\circ}, 270^{\circ}$). By generating orthogonal side views, the pipeline can directly acquire pixel data for the lateral torso and underarms. To resolve ambiguities in overlapping regions, a training-free confidence-aware fusion module can be employed, where the final color $C(\mathbf{P})$ for a surface point $\mathbf{P}$ is determined by a weighted average: $C(\mathbf{P}) = (\sum w_i C_i) / \sum w_i$, using the geometric confidence $w_i = \max(0, \mathbf{n} \cdot \mathbf{v}_i)$. This weighting ensures that texels are sourced from the camera with the most direct line-of-sight, effectively voting out tangent distortions. 

Furthermore, while ImageReward provides a high-fidelity proxy for human preference, we plan to conduct a large-scale user study involving professional digital artists to further validate the production-readiness of our synthesized assets.

Finally, we acknowledge the rapid evolution of large language and multimodal models. While new versions such as GPT-5 or specialized variants are released frequently, benchmarking against every emerging model is impractical due to the pace of development. However, our methodology is fundamentally model-agnostic. The current implementation utilizes GPT-4o as the state-of-the-art generative prior at the time of study; nevertheless, the core architecture can seamlessly integrate future models as they become available. This ensures the long-term relevance of our contribution even as generative world models continue to advance.

\section{Conclusion}
In this study, we proposed SMPL-GPTexture, a training-free framework that effectively generates high-quality, photorealistic textures for 3D human avatars. Our tailored prompts and geometry-aware inverted rasterization ensure that the generated texture maps are both accurate and complete. Experimental results demonstrate that our method achieves excellent geometric coherence and texture fidelity compared with state-of-the-art methods. Beyond strong performance, SMPL‑GPTexture offers high generalizability: as new world models emerge, our framework can seamlessly adopt them without retraining. Its prompt-driven design enables virtually infinite texture diversity, making it well-suited for open-ended applications such as digital fashion, stylized avatars, and synthetic data generation.

\bibliography{references}
\end{document}